\documentclass[11pt,a4paper]{article}
\pdfoutput=1
\usepackage{empheq}
\usepackage{cases}
\usepackage{a4wide}
\usepackage{amsfonts}
\usepackage{amssymb}
\usepackage{amsmath}
\usepackage{graphicx}
\usepackage{booktabs}
\usepackage{array}
\usepackage{color}
\usepackage{ulem}
\usepackage[small,bf]{caption}
\usepackage{cite}
\usepackage[usenames,dvipsnames]{xcolor}
\usepackage{subfigure}
\usepackage{verbatim}
\usepackage[symbol]{footmisc}

\allowdisplaybreaks

\numberwithin{equation}{section}

\newcounter{mysubequation}[equation]

\begin{document}
\begin{titlepage}

\begin{center}
{
\bf\LARGE Dark Photon-Photon Resonance Conversion\\[0.3em]
of GRB221009A through Extra Dimension
}
\\[8mm]
M.~Afif~Ismail,\textsuperscript{1}\footnote[1]{m.afif.ismail@gmail.com} Chrisna~Setyo~Nugroho,\textsuperscript{1}
\footnote[2]{setyo13nugros@ntnu.edu.tw}
 and Qidir Maulana Binu
Soesanto\textsuperscript{2}\footnote[3]{ qidirbinu@fisika.fsm.undip.ac.id} 
\\[1mm]
\end{center}
\vspace*{0.50cm}

\centerline{\it \textsuperscript{1} Department of Physics, National Taiwan Normal University, Taipei 116, Taiwan}
\centerline{\it \textsuperscript{2} Department of Physics, Faculty of Sciences and Mathematics,}
\centerline{\it Universitas Diponegoro, Jl. Prof. Jacob Rais, Semarang 50275, Indonesia}
\vspace*{1.20cm}

\begin{abstract}
\noindent
The recently observed very high energy (VHE) photons dubbed as
GRB221009A by several terrestrial observatories such as LHAASO
and Carpet-2 require a physics explanation beyond the standard model.
Such energetic gamma ray bursts, originating from yet unknown  very
distance source at redshift z = 0.1505, would be directly
scattererd by
extragalactic background lights (EBL) rendering its improbable detection at the earth. We show that dark photon which resides
in extra dimension would be able to resolve this issue when it
oscillates resonantly with the photon in similar fashion with 
neutrino oscillation. We demonstrate that, for dark photon
mass equals to 1 eV, the probability of GRB221009A photons with energy above 0.2 TeV to reach the earth lies in the range
between $10^{-6} \, \text{to} \, 95\% $ for kinetic mixing
values $10^{-14} \leq \epsilon \leq 10^{-12.5}$ allowed by current  constraints.   
\end{abstract}

\end{titlepage}
\setcounter{footnote}{0}

\section{Introduction}

On October $9^{\text{th}}$ 2022, several observatories such as LHAASO \cite{Y.Huang,LHAASO:2023kyg}, Carpet-2 \cite{D.Dzhappuev}, and Fermi GBM~\cite{P.Veres}, observed extra-energetic
gamma-ray burst at redshift z = 0.1505 or approximately $636$ Mpc. This anomalous high energy
photon, called GRB221009A~\cite{S.Dichiara}, was detected with energy above 0.2
TeV, reaching even higher energy range between 10 TeV to 18
TeV as reported by LHAASO. Moreover, Carpet-2 recorded GRB221009A photon energy up to 251 TeV coming from yet unknown
source. Such ultra-energetic photons coming from very distant
source are known to be converted by EBL into electron-positron
pairs~\cite{Gould:1966pza,Fazio:1970pr}. As a result, these photons would be suppressed effectively before reaching the earth. 

This phenomena, which can not be explained by the standard
model (SM) of particle physics, calls for a new explanation
beyond the SM. Many beyond standard model (BSM) hypotheses has
been proposed to resolve this issue including the photon to
axion-like particle (ALP) conversion~\cite{Zhang:2022zbm,Lin:2022ocj,Galanti:2022chk,Baktash:2022gnf,Troitsky:2022xso,Nakagawa:2022wwm,Aban:2023amc,Wang:2023okw}, high energy photon
induced by heavy neutral lepton decay~\cite{Huang:2022udc,Cheung:2022luv,Brdar:2022rhc,Guo:2023bpo}, as well as the violation of Lorentz invariance by cosmic high energy photon~\cite{Li:2022wxc,Li:2022vgq,Finke:2022swf,Zhu:2022usw,Li:2023rgc}.     

In this paper, we propose photon-dark photon oscillation to
explain GRB221009A. Dark photon is well studied theory of BSM. However, as photon and dark photon (DP)
mixes kinetically by the mixing angle $\epsilon$, one expects
that the conversion probability of photon into DP is
suppressed by $\epsilon^{2}$. Based on the current stringent constraints of $\epsilon$
from both experiments and observations~\cite{Workman:2022ynf,Fabbrichesi:2020wbt,Caputo:2021eaa,Hook:2021ous,Ismail:2022ukp,Carenza:2023qxh,Chen:2024jbr,Kalashev:2018bra}, this conversion is strongly attenuated. This problem can be alleviated when dark photon resides in extra dimension. The
notion that dark photon as well as dark matter (DM) propagate
in the bulk of extra dimension was proposed in~\cite{Rizzo:2018ntg,Rizzo:2018joy,Rizzo:2020ybl}. Actually,
the theory of extra dimension is well known theory to address
several problems like unification theory, hierarchy problem in
particle physics, the origin of neutrino masses, as well as recent particle physics anomalies~\cite{Deffayet:2000pr,Arkani-Hamed:1998jmv,Antoniadis:1998ig,Dienes:1999gw,Aban:2021lop,Aban:2023pgq}. 
In typical extra dimensional theory, any exotic particles like  right-handed neutrino, ALP, dark photon, as well as graviton
reside in higher dimensional space called bulk. On the other hand, the SM particles are constrained to live in 4 dimensions
denoted as the brane. The particles that live in the bulk are allowed to
propagate into the brane while the SM particles always stay on the brane. 

As the origin of GRB221009A is yet to be known, we do not discuss the
specific model of the source. Instead, we assume that the high
energy photon is converted to dark photon when it propagates
in the intergalactic medium and converted back into photon
when the dark photon arrives at the Milky Way galaxy. As dark photon can propagate through the bulk and brane, the effect of
extra dimension has to be taken into account during its propagation. We show that the effect of extra dimension
induces the resonance conversion probability which effectively cancel the $\epsilon^{2}$ suppression allowing GRB221009A gamma ray burst to reach the earth.

The rest of this paper is organized as follows. In
Section~\ref{sec:the model}, we discuss photon-dark photon interaction as well as its conversion probability via extra
dimension. We show our numerical results together with its
implications in Section~\ref{sec:discussion}. Our summary and conclusions are presented in Section~\ref{sec:Summary}.

\section{Photon-Dark Photon Conversion}
\label{sec:the model}

Dark photon is a remnant of hidden $U(1)^{'}$ symmetry
or other non-Abelian symmetries which is spontaneously broken
ala Higgs mechanism~\cite{Holdom:1985ag,Okun:1982xi,Fayet:1980rr,Georgi:1983sy,Huang:2015wts,Dirgantara:2020lqy,Chen:2019pnt,Tran:2023lzv}. Moreover, its mass can also be generated explicitly via Stuckelberg mechanism. As an extension of the standard model, dark photon is well
known for its direct mixing with the photon. This mixing is generated at dimension 4 operator without any higher dimensional scale suppression. The Lagrangian encoding this interaction is given by
\begin{align}
	\mathcal{L} \supset -\frac{1}{4}F^{\mu\nu}F_{\mu\nu}-\frac{1}{4}F^{'\mu\nu}F^{'}_{\mu\nu}+\frac{\epsilon}{2}F^{\mu\nu}F^{'}_{\mu\nu}+\frac{1}{2}m_{\gamma'}A^{'\mu}A^{'}_{\mu}\,.
 \label{lagrangian}
\end{align}
Here, $F^{\mu\nu}$ ($F'^{\mu\nu}$) corresponds to the field
strength tensor of the photon field $A_{\mu}$ (dark photon
field $A'_{\mu}$). The third term captures the kinetic
mixing between photon and dark photon with mixing parameter
 $\epsilon$ while the last term gives dark photon mass.

In dealing with photon-dark photon oscillation, it is more convenient to write the Lagrangian in active-sterile basis analogue to that of neutrino oscillation~\cite{Brahma:2023zcw} 
\begin{align}
\label{eq:DPbasis}
	\mathbf{A}^{\mu}&\equiv
	\begin{pmatrix}
		A^{\mu}_{a} \\
		A^{\mu}_{s}
	\end{pmatrix}
	=
	\begin{pmatrix}
		1 & 0 \\
		-\epsilon & 1 
	\end{pmatrix} 
        \begin{pmatrix}
		A^{\mu} \\
		A^{'\mu}
	\end{pmatrix}
	+ \mathcal{O}(\epsilon^{2})\,,
\end{align}
where, after expanding up to the leading order of $\epsilon$,  $A^{\mu}_{a}$ and $A^{\mu}_{s}$ denote the active and sterile fields, respectively.
In this new basis, the Lagrangian in Eq.~(\ref{lagrangian}) can be written as
\begin{align}
	\mathcal{L} = -\frac{1}{4}F_{a}^{\mu\nu}F^{a}_{\mu\nu}-\frac{1}{4}F_{s}^{\mu\nu}F^{s}_{\mu\nu}+\frac{1}{2} \mathbf{A}^{T}_{\mu} \mathcal{M}^{2}\mathbf{A}^{\mu} + \mathcal{O}(\epsilon^{2})\,.
 \label{eq:lagrangian_new}
\end{align}
Here, $F_{a}^{\mu\nu}$ and $F_{s}^{\mu\nu}$ correspond to the field strength tensors of the active and sterile field, respectively. Moreover, the SM particles interact with the active state with coupling strength $e$ i.e. $e\,J_{\mu} A^{\mu}_{a}$.
The last term of eq.\eqref{eq:lagrangian_new} gives the mass-squared mixing matrix
\begin{align}
\label{eq:matrixmixing}
	\mathcal{M}^{2}& \approx
	\begin{pmatrix}
		m^{2}_{\text{eff}}&\epsilon\, m_{\gamma'}^{2} \\
		\epsilon\, m_{\gamma'}^{2} & m_{\gamma'}^{2}
	\end{pmatrix}\,,
\end{align}
which captures the mixing between $A^{\mu}_{a}$ and $A^{\mu}_{s}$. The square of photon effective mass $m^{2}_{\text{eff}}$ appearing in the first diagonal element of eq.\eqref{eq:matrixmixing} signifies the medium effect which
takes zero value if the photon propagates in
vacuum\footnote{Here, we assume that $m^{2}_{\text{eff}}$ is
constant which implies that the variation of $m^{2}_{\text{eff}}$ is negligible compared to the wave number $k^{-1}$ of $\mathbf{A}^{\mu}$~\cite{Brahma:2023zcw}.}. 

To compute the conversion probability of photon to dark photon
or vice versa, one needs to solve the Klein-Gordon (KG)
equation corresponds to the Lagrangian given by eq.\eqref{eq:lagrangian_new}, which in the frequency domain reads  
\begin{align}
	(\omega^{2}-k^{2}-\mathcal{M}^{2})\,\mathbf{\tilde{A}}^{\mu}(\omega,k)\,=\,0\,,
        \label{eq:eom-tm}
\end{align}
where $\mathbf{\tilde{A}}^{\mu}(\omega,k)$ is the corresponding Fourier transform of $\mathbf{A}^{\mu}$. In   relativistic limit relevant for GRB221009A, $\omega \approx k \gg m_{\gamma^{'}}, m_{\text{eff}}$ one arrives at linearized Schrodinger-like equation~\cite{Raffelt:1987im}
\begin{align}
	i\partial_{z}\mathbf{A}^{\mu}=H_{0}\mathbf{A}^{\mu}\,,
 \label{eq:Scheom}
\end{align}
where the explicit form of the Hamiltonian is given by
\begin{align}
	H_{0}&=
	\begin{pmatrix}
		\omega + \Delta_\text{pl} & \epsilon\, \Delta_{A'} \\
		\epsilon\, \Delta_{A'} & \omega + \Delta_{A'}
	\end{pmatrix}\,.	
\end{align}
Here, we replace the time derivative with the spatial one (at $z$ direction) since we are dealing with relativistic particle. The explicit expressions of $\Delta_{\text{pl}}$ and $\Delta_{A^{'}}$ are
\begin{align}
\label{eq:deltas}
\Delta_{\text{pl}} \,=\, -\frac{m^{2}_{\text{eff}}}{2\,E}\,\,\, \text{and}\,\,\, \Delta_{A^{'}} \,=\, - \frac{m^{2}_{\gamma^{'}}}{2\,E}\,,
\end{align}
where we have substituted $\omega$ with $E$. 

Up to this point, we have not discussed the implication of dark
photon propagation in extra dimension. As dark photon may
traverse via the bulk in higher dimensional space as well as the brane
in four dimension, one needs to take this effect into account. We assume that the ultra high energy GRB221009A photons are converted to dark photons during their propagation in the
intergalactic medium. Furthermore, their reconversion into photons occurs when dark photons arrive at the edge of the
Milky Way galaxy. Finally, these reconverted photons propagates through the galactic medium before reaching the earth.

Dark photon may traverse into two different paths prior to its arrival at the border of the Milky Way galaxy. The first path is
through the usual four dimensional spacetime or the brane. In this case, dark photon travels directly to the edge of the Milky Way galaxy
without experiencing any attenuation thanks to its elusive nature. On the other hand, it may travel through the bulk before reappear on the brane located at the Milky Way border.    Since the brane is embedded in higher dimensional space, there would be a path difference in dark photon propagation.
This difference would be zero if the brane were flat and rigid in its embedding. On the other and, if the embedding is curved, the path differrence~\cite{Pas:2005rb} 
\begin{align}
\label{eq:zdiff}
\delta \,=\, \frac{z_{\text{bulk}}-z_{\text{brane}}}{z_{\text{bulk}}}\,,
\end{align} 
would be different from zero. This effect gives an additional term in the Hamiltonian~\cite{Pas:2005rb}
\begin{align}
\label{eq:Hdelta}
H_{\delta} \,=\, \begin{pmatrix}
		\Delta_\delta & 0 \\
		0 & -\Delta_\delta
	\end{pmatrix} \,\,\, \text{with}\,\, \Delta_{\delta} \,=\, -\frac{E\,\delta}{2}\,.
\end{align}
Taking into account this additional term, the linearized Schrodinger-like equation becomes
\begin{align}
\label{eq:Schfin}
i\partial_{z}\mathbf{A}^{\mu}&=(H_{0}+H_{\delta})\,\mathbf{A}^{\mu} \equiv H\,\mathbf{A}^{\mu}\,,
\end{align}
\begin{align}
\label{eq:Htot}
H &= \begin{pmatrix}
	E + \Delta_\text{pl} +\Delta_{\delta} & \epsilon\, \Delta_{A'} \\
	\epsilon\, \Delta_{A'} & E + \Delta_{A'} -\Delta_{\delta}
	\end{pmatrix}
\end{align}

To solve eq.\eqref{eq:Schfin}, one needs to diagonalize the Hamiltonian $H$ 
\begin{align}
\label{eq:DiagH}
	V H V^\dagger &= 
	\begin{pmatrix}
		\lambda_1 & 0 \\
		0 & \lambda_2 
	\end{pmatrix} \,,
\end{align}
with the corresponding $V$ matrix as well as its explicit elements given by
\begin{align} 
\label{eq:Vmatrix}
	V &= \begin{pmatrix}
		\cos\theta & -\sin\theta \\
		\sin\theta & \cos\theta 
	\end{pmatrix} \,\,\,\text{with}\,\,\tan2\theta 
	= \frac{2\epsilon \Delta_{A'}}{\Delta_{A'}-\Delta_\text{pl} - 2 \Delta_{\delta}}\,.
\end{align}
The corresponding eigenvalues of the Hamiltonian are
\begin{align}
\label{eq:eigenvalues}
	\lambda_{1,2}
	= \frac{1}{2}\left\{(2\,E + \Delta_\text{pl} + \Delta_{A'})\pm \sqrt{4\epsilon^2 \Delta_{A'}^2 + (\Delta_\text{pl} - \Delta_{A'}+2\Delta_{\delta})^2}\right\} .
\end{align} 
In this diagonal basis, $\tilde{\mathbf{A}}^{\mu}=V\mathbf{A}^{\mu}$ the propagation of the field $\tilde{A}^{\mu}_{j}$ from initial point at $z_{0}$ to $z$ is 
\begin{align} 
	\tilde{A}^{\mu}_{j}(z)=e^{-i\,(z-z_0)\,\lambda_j}\,\tilde{A}^{\mu}_{j}(z_0)\,.
\end{align}
To calculate the conversion probability from active state to the sterile state, one simply invert back the relation $\tilde{\mathbf{A}}^{\mu}=V\mathbf{A}^{\mu}$ into $\mathbf{A}^{\mu}=V^{\dag}\tilde{\mathbf{A}}^{\mu}$. 
Assuming that we have a photon (active field) in the initial state at $z_{0}=0$, the probability of detecting dark photon (sterile field) after traversing a distance $z$ is
\begin{align}
\label{eq:probAtoS}
	P_{\gamma \rightarrow \gamma^{'}} = \left|\langle A_{s}(z)|A_{a}(z_0)\rangle\right|^2\,,
\end{align}
\begin{align}
\label{eq:probAtoSEd}
	P_{\gamma \rightarrow \gamma^{'}} 	&= \frac{4\epsilon^2 \Delta_{A'}^2 \sin^2\left(\frac{z}{2}\sqrt{4\epsilon^2 \Delta_{A'}^2 + (\Delta_\text{pl} - \Delta_{A'}+2\Delta_{\delta})^2}\right)}{4\epsilon^2 \Delta_{A'}^2 + (\Delta_\text{pl} - \Delta_{A'} + 2\Delta_{\delta})^2}\,.
\end{align}

As one can see in eq.~\eqref{eq:probAtoSEd}, the photon to dark
photon conversion probability is suppressed by $\epsilon^{2}$. Consequently, as the current experimental as well as
observational limits of $\epsilon$ are quite severe, it is not
possible for GRB221009A to reach the earth. However, this is
not the case when the resonant condition is satisfied. At resonance, the diagonal elements of the Hamiltonian in eq.~\eqref{eq:Htot} have the same value, implying the following relation to hold
\begin{equation}
\Delta_{A'} - \Delta_{\text{pl}} = 2\Delta_{\delta}\,. 
\label{eq:resonance}
\end{equation}
At resonance, the conversion probability becomes 
\begin{align}
\label{eq:ProbRes}
	P^{\text{res}}_{\gamma \rightarrow \gamma^{'}} = \sin^2\left(z\, \epsilon\,\Delta_{A^{'}}\right)\,. 
\end{align}
Here, the $\epsilon$ only appears inside the sine argument.
Furthermore, we can extract the resonant energy as
\begin{align}
\label{eq:ERes}
E_{\text{res}} \approx \sqrt{\frac{m^{2}_{\gamma^{'}}}{2\,\delta}}\,,
\end{align}
where we have used the fact that $\Delta_{A^{'}} \gg \Delta_{\text{pl}}$ as we will demonstrate in the following section. This resonance energy may have different values 
depending on the variation of $\delta$ for a fixed dark photon mass. We will employ the resonant conversion probability c.f. eq.~\eqref{eq:ProbRes} in the rest of our paper.

\section{Results and Discussion}
\label{sec:discussion}

To demonstrate the ability of gamma ray burst GRB221009A to
reach the earth after its conversion into dark photon, we set
the dark photon mass to 1 eV. Moreover, we take four benchmark
values of $\epsilon$ for the corresponding dark photon mass  allowed by current limits~\cite{Cline:2024qzv}: $\epsilon = 10^{-12.5},\,10^{-13},\,10^{-13.5},\,\text{and}\,10^{-14}$.    
\begin{figure}[h!]
	\begin{center}
		\includegraphics[width=0.7\textwidth]{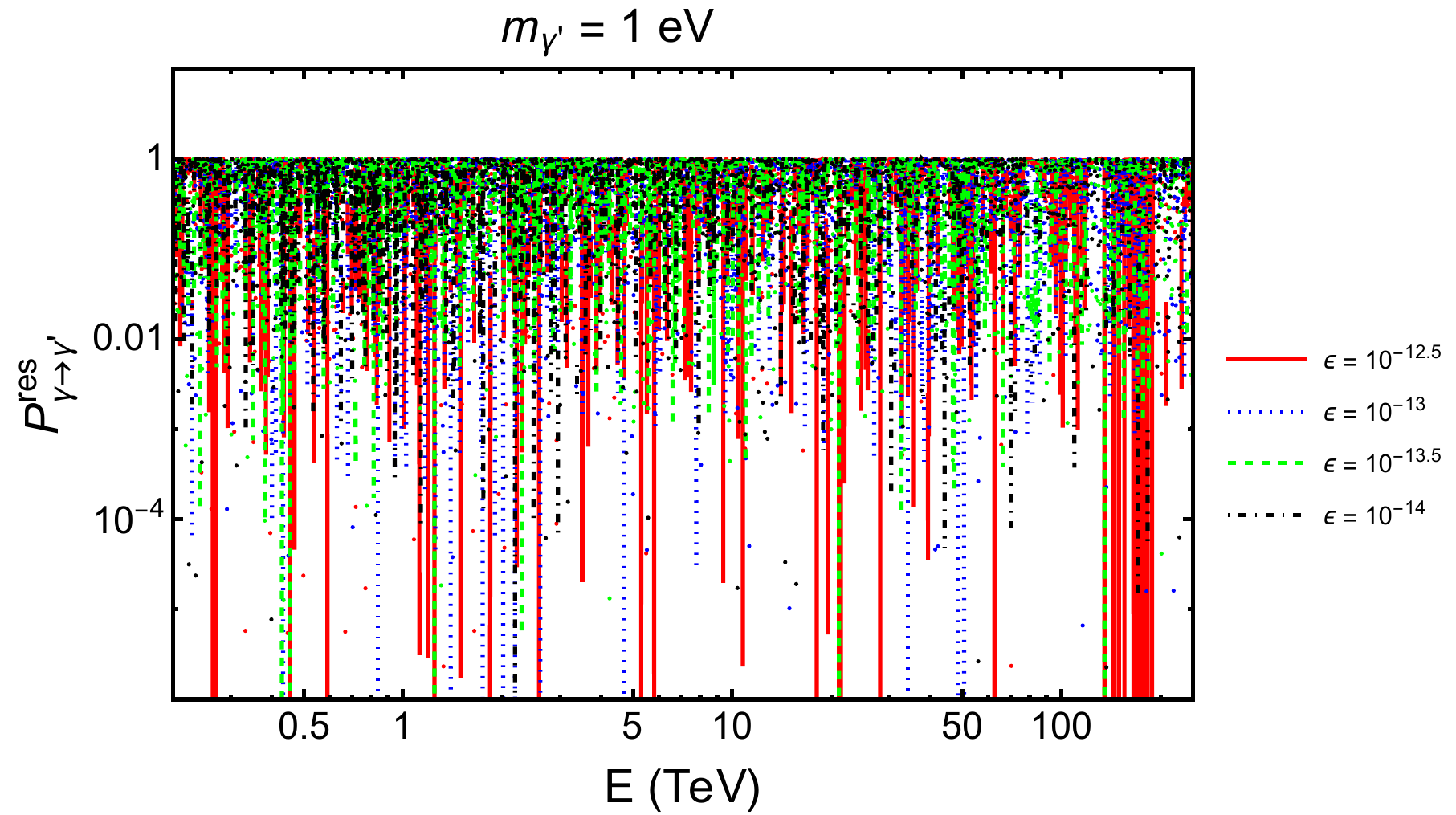}
		\caption{Resonance conversion of photon into dark photon in the intergelactic medium for different $\epsilon$: $10^{-12.5}\,(\text{red solid line}),\, 10^{-13}\,(\text{blue dotted line}),\,10^{-13.5}\,(\text{green dashed line})$, and $10^{-14}$ (black dot-dashed line). We set dark photon mass to 1 eV.}
		\label{fig:PAtoS}
	\end{center}
\end{figure}

\begin{figure}[h!]
	\centering
	\subfigure[Higher energy regime of fig.\ref{fig:PAtoS}]{%
		\includegraphics[width=0.4\textwidth]{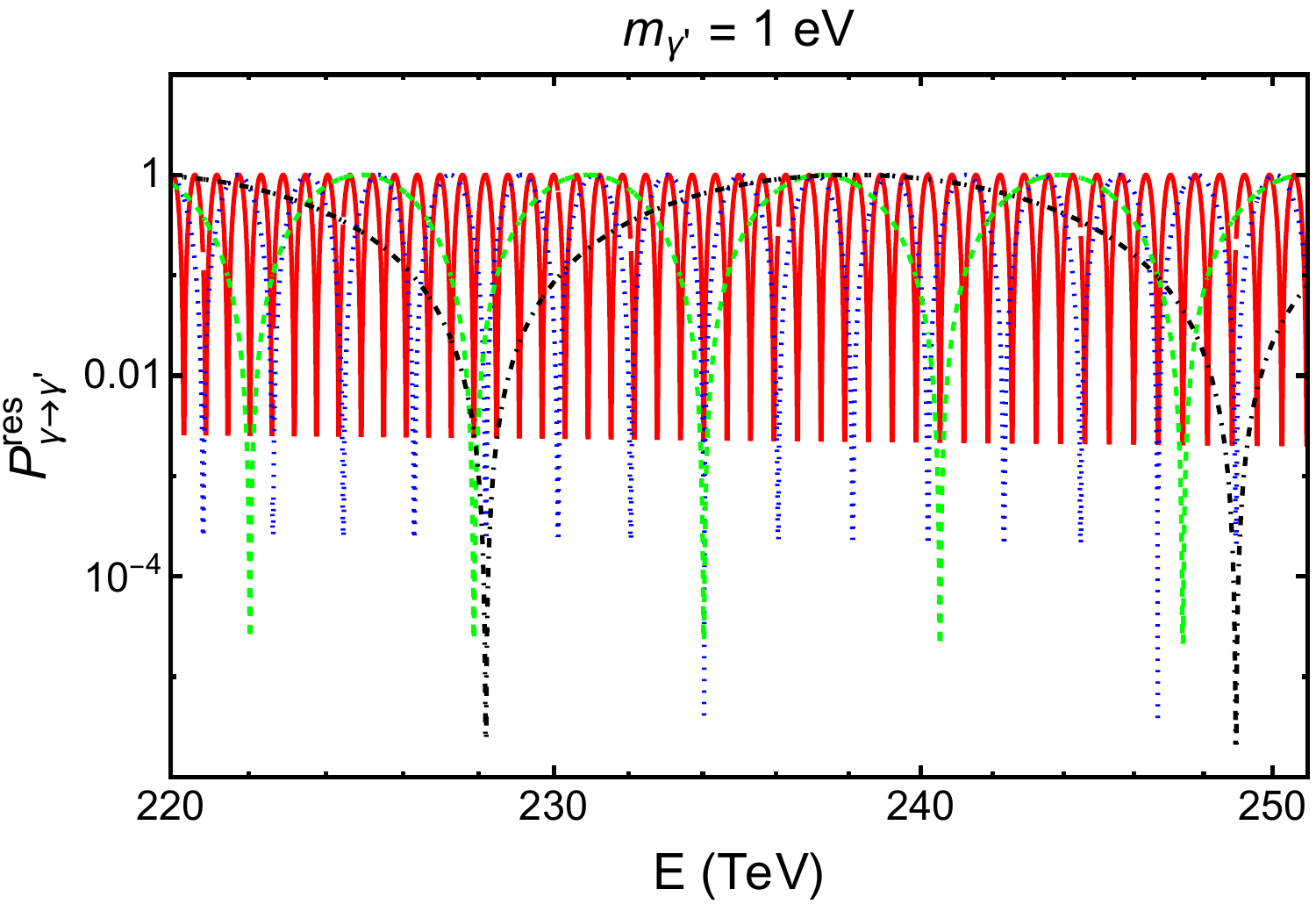}
	}\quad
	\subfigure[Lower energy regime of fig.\ref{fig:PAtoS}]{%
		\includegraphics[width=0.5\textwidth]{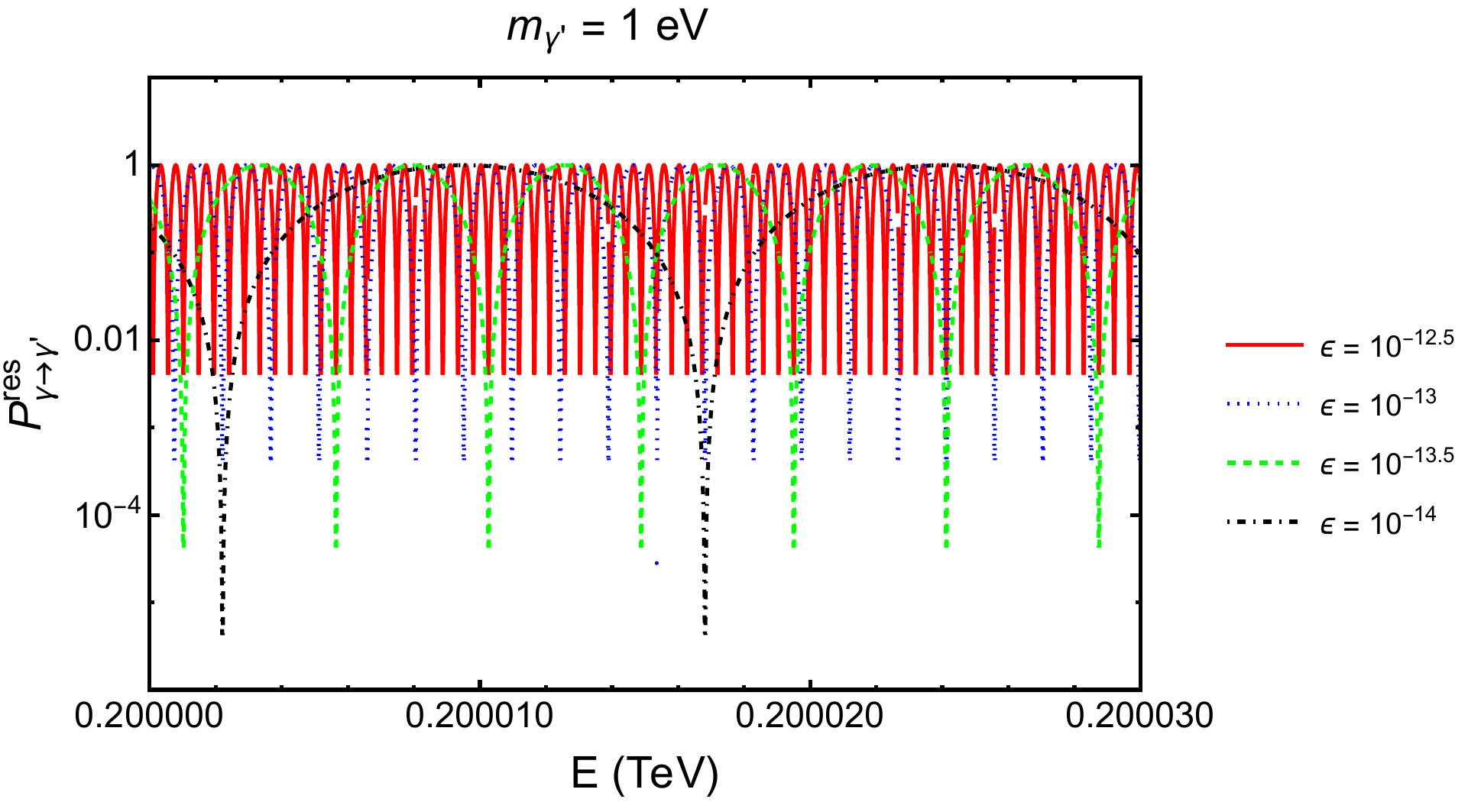}
	}
	\caption{Higher and lower energy regime of photon to dark photon conversion. The squared sinusoidal behaviour is visible in both figures.}
	\label{fig:PAtoSZoomed}
\end{figure}

In fig.\ref{fig:PAtoS}, we show the resonant conversion
probability of the GRB221009A photon into dark photon as a function of photon energy
in the intergalactic medium. There, the photon energy is
taken from 0.2 GeV as observed by LHAASO to 251 GeV as reported by Carpet-2 collaboration. It displays a very fast
oscillation with conversion probability ranging from 0 to 1 as
expected from eq.\eqref{eq:ProbRes}. The origin of this abrupt
oscillation is due to the large value of the path $z$
traversed by the dark photon i.e. the distance from the redshift z = 0.1505 (or approximately 636 Mpc) to the edge of the Milky Way  galaxy ($\approx 23.95$ kpc~\cite{Wang:2023okw}).
We take the value
of $m_{\text{eff}} \approx 2\times 10^{-14}$ eV relevant for
the typical intergalactic medium~\cite{Brahma:2023zcw}. Consequently, since $m_{\gamma^{'}} \gg m_{\text{eff}}$ which implies $\Delta_{A^{'}} \gg \Delta_{\text{pl}}$ (see eq.\eqref{eq:deltas}), the resonance energy
approximation given by eq.\eqref{eq:ERes} is valid. To show the square of sinusoidal characteristic of the $P^{\text{res}}_{\gamma\rightarrow \gamma^{'}}$, we plot the zoomed in version
of fig.\ref{fig:PAtoS} in fig.\ref{fig:PAtoSZoomed} for the
higher energy regime (left-panel) as well as the lower energy
regime (right-panel). In these two panels, the square
sinusoidal behavior is apparent, with the slower oscillation
behaviour as the value of the mixing parameter $\epsilon$ is reduced. Following~\cite{Aban:2023amc}, we set $P^{\text{res}}_{\gamma \rightarrow \gamma^{'}} = 95 \%$ to optimize the number of photon being converted to dark photon.    
\begin{figure}[h!]
	\begin{center}
		\includegraphics[width=0.7\textwidth]{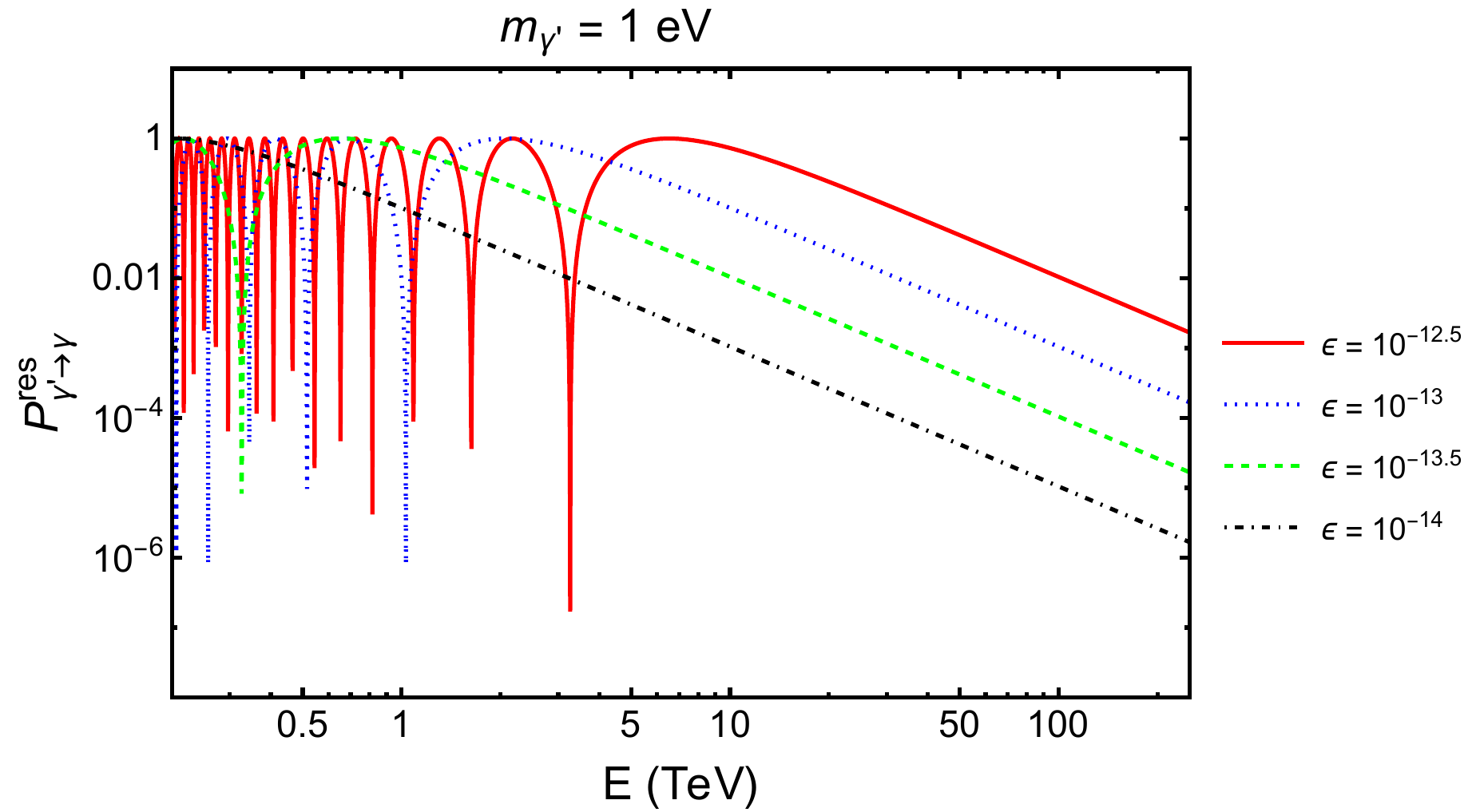}
		\caption{Resonance conversion of dark photon into photon at Milky Way galaxy for different $\epsilon$: $10^{-12.5}\,(\text{red solid line}),\, 10^{-13}\,(\text{blue dotted line}),\,10^{-13.5}\,(\text{green dashed line})$, and $10^{-14}$ (black dot-dashed line). }
		\label{fig:PStoA}
	\end{center}
\end{figure}

The reconversion probability of dark photon into photon at the border of Milky Way galaxy is displayed in fig.\ref{fig:PStoA}.
As the electron density is higher inside the Milky Way disk $n_{e} \approx 1.1 \times 10^{-3} \text{cm}^{-3}$, the effective mass of the photon $m^{2}_{\text{eff}} = 4 \pi \alpha n_{e}/m_{e}$ is approximately $4.1 \times 10^{-12}$ eV~\cite{Wang:2023okw}.
Here, $\alpha$ and $m_{e}$ denote the fine structure constant and electron mass, respectively. Moreover, we take the
propagation distance from the edge of the Milky Way to the earth equals to $z = 23.95$ kpc~\cite{Wang:2023okw}.   
We see from fig.\ref{fig:PStoA} that the oscillation becomes
visible as the propagation length gets smaller.
Furthermore, despite it still varies between 0 to 1, the
reconversion probability decays as the energy increases. This
can be understood from the argument of sine in
eq.~\eqref{eq:ProbRes} which proportional to $\Delta_{A^{'}}$ c.f. eq~\eqref{eq:deltas}. The total probability of gamma ray
burst GRB221009A arriving at the earth is obtained by the
product of $P^{\text{res}}_{\gamma \rightarrow \gamma^{'}}$
and $P^{\text{res}}_{\gamma^{'} \rightarrow \gamma}$. Using
four benchmark values of $\epsilon$, the total probability ranges between
$10^{-6}$ to $95\%$ which indicates that ultra-energetic photons from GRB221009A can reach the earth without any significant suppression.   

As an additional remark, let us comment on the implication of gamma ray burst GRB221009A for the brane fluctuation. 
In the resonant energy range of GRB221009A from 0.2 TeV to 251 TeV, the corresponding path difference parameter $\delta$ lies between $1.25 \times 10^{-23} \geq \delta \geq 7.94 \times 10^{-30}$. The
brane fluctuation associated with this range, which equals
to the square-root of $\delta$~\cite{Pas:2005rb}, varies within the interval $3.54 \times 10^{-12} - 2.82 \times 10^{-15}$. This is more
stringent than the limit extracted from the neutrino
oscillation experiment such as LSND/KARMEN, which sets the value of the brane
fluctuation to range over $10^{-8} - 10^{-9}$~\cite{Pas:2005rb}.   

\section{Summary and Conclusions}
\label{sec:Summary}

The arrival of ultra-high energy photons known as GRB221009A
at the earth, as observed by terrestrial observatories such as
LHAASO and Carpet-2, calls a physics explanation beyond the
SM. We propose a novel idea using photon-dark photon
conversion through extra dimension to explain this phenomena. We show that, if dark photon resides in the higher dimensional
space, the photon-dark photon conversion probability would attain resonant enhancement allowing the GRB221009A gamma ray
burst to be detected at the earth. By setting dark photon mass
to 1 eV as well as taking four benchmark values of the photon-dark photon mixing parameter $\epsilon = 10^{-12.5},\, 10^{-13},\, 10^{-13.5}\,,\text{and}\,10^{-14}$, we demonstrate
that there is $10^{-6}$ up to $95\%$ fraction of the initial
GRB221009A photon flux
would be detectable at the terrestrial gamma ray observatories. Furthermore, this implies much
stronger constraint on the brane fluctuation with respect to
the limit extracted from neutrino oscillation experiment.

\section*{Acknowledgment}
\label{sec:Acknowledgment}
CSN is supported by the National
Science and Technology Council (NSTC) of Taiwan under Grant No. NSTC
113-2811-M-003-019.

\end{document}